\documentclass[11pt]{article}
\usepackage{amsmath,amssymb}
\begin{document}
\oddsidemargin .03in
\evensidemargin 0 true pt
\topmargin -.4in
 
%Abbreviations %
\def\ra{{\rightarrow}}
\def\a{{\alpha}}
\def\b{{\beta}}
\def\l{{\lambda}}
\def\eps{{\epsilon}}
\def\pr{{\partial}}
\def\tri{{\triangle}}
\def\na{{\nabla }}
\def\sp{\vspace{.15in}}
\def\hs{\hspace{.25in}}
\def\n{\nonumber}
\def\ni{{\noindent}}
\def\Ra{{\Rightarrow}}

\newcommand{\be}{\begin{equation}} \newcommand{\ee}{\end{equation}}
\newcommand{\bea}{\begin{eqnarray}}\newcommand{\eea}{\end{eqnarray}}

%********************************************************************%

\begin{titlepage}
\topmargin= -.2in
\textheight 9.5in

\begin{center}
\baselineskip= 18 truept

\vspace{.3in}

\centerline{\Large\bf  Deformation of $D_p$-Brane Worldvolume in Weakly Curved Background}

\vspace{.6in}
\noindent
{\bf Richa Kapoor}\footnote{richa.phy@gmail.com }

\vspace{.2in}

\noindent

\noindent
{{\Large Department of Physics \& Astrophysics}\\
{\Large University of Delhi, New Delhi 110 007, India}}

\vspace{.2in}

{\today}
\thispagestyle{empty}

\vspace{.6in}
\begin{abstract}

\baselineskip=14 truept  

\vspace{.12in}
We study a $D_p$-brane in a parallelizable NS-NS background. The article starts with a brief review of the non-associative deformation of $D$-brane worldvolume in presence of torsion \cite{cornalba}. We suggest an alternative form and heuristic derivation of the open string metric for weakly curved backgrounds, by promoting the  constant two-form in the flat space formula to a dynamical two-form and then Taylor expanding the bulk fields in Riemann Normal Coordinates at the origin. For weakly curved backgrounds, terms only upto the leading order in the NS-NS field strength or torsion contribute. This formalism differs from the author's earlier works in a collaboration \cite{richa}. We use the open string metric proposed in this paper to determine the deformation of $D_5$-brane for a particular NS-NS background. It turns out that a spherical $D_5$-brane with torsion acts like an extremal black 5-brane in the limit of radius $r \rightarrow 0$. This is an interesting case of gauge/gravity duality with an intractable non-commutative and non-associative gauge theory on a $D_p$-brane with torsion, which is equivalent to a simpler and ordinary gravity theory on the extremal black brane.

\vspace{1in}

\noindent

\noindent

\end{abstract}
\end{center}

\vspace{.2in}

\baselineskip= 16 truept

\vspace{1in}

\end{titlepage}

\baselineskip= 18 truept
   
\section{Introduction}    
Since the paper \cite{seiberg} by Seiberg and Witten, there has been an increased interest in the study of $D_p$-branes in type II superstring theory with non-vanishing NS-NS $B$-field. The main feature in presence of the $B$-field is that the $D$-brane worldvolume becomes non-commutative. With T-duality, the closed string theories treat NS-NS two-form and dilaton on the same footing as the metric. Thus string theory admits non-geometric backgrounds also as consistent solutions.  An important quantity in the Born-Infeld action is the gauge invariant combination ${\cal F} = B + F$.  When a $D_p$-brane is placed in a background with non-zero $B$-field, three possibilities arise \cite{cornalba, seiberg, schomerus, ho-yeh, burban}:
\begin{enumerate}       
 \item ${\cal F}$ is constant with vanishing Maxwell field strength, $F = dA = 0$. This case is very well understood and it corresponds to a flat brane in a flat spacetime with constant $B$-field background. Here we have a noncommutative and associative Moyal deformation of the brane worldvolume.
 \item ${\cal F}$ is not constant but $d{\cal F} = 0$. In this case ${\cal F}(x) = B + F(x)$ i.e., $B$ is constant hence $dB = 0$. Also, $dF = d^2A = 0$ as $d^2 = 0$. This case is an extension of Moyal deformation, to include varying symplectic two-form ${\cal F}$. It leads to the noncommutative and associative Kontsevich deformation of the brane worldvolume. It corresponds to a curved $D$-brane in a flat background.
 % i.e., $\partial_\mu{\cal F}_{\nu\rho} + \partial_\nu{\cal F}_{\rho\mu} + \partial_\rho{\cal F}_{\mu\nu} = 0$;  
 \item The general case is of a curved $D$-brane in a curved background of NS-NS three form flux. Here,  torsion $H = d{\cal F} = dB + dF  = dB \neq 0$. Since ${\cal F}$ is not closed, so it is not a symplectic form. The ordinary product of algebra on a $D_p$-brane is replaced by a noncommutative and nonassociative Kontsevich star product.  
\end{enumerate}       
            
\sp\ni  
This article deals with the general case, which is briefly reviewed in the section \ref{sec1} and \ref{sec2}, mainly following the Ref. \cite{cornalba}. In presence of torsion, the coordinates and momenta of the open string endpoints satisfy mixed commutation relations. This leads to a new type of noncommutative spacetime  which is also nonassociative. For such braneworld scenarios incorporating torsion, two well known techniques are often used - the sum rule which gives information about the brane torsion terms, and the Taylor expansion which describes the bulk torsion terms. The background fields are Taylor expanded with the derivative terms interpreted as new interactions. This enables us to determine the open string metric $G$ and the deformation $\theta$. Section \ref{sec3} suggests an alternative heuristic derivation  of the open string metric in a weakly curved background. This form is the author's  independent idea, and is different from the open string metric given in her earlier works in a collaboration \cite{richa}. In Section \ref{sec4}, we choose a particular $B$-field solution and use it to compute the emergent geometry on a $D_5$-brane. It is evident that a flat $D_p$-brane with torsion pulled back or projected onto its  worldvolume, equals a curved $D_p$-brane in the curved background.  

\section{Parallelizable manifolds}   \label{sec1}  
A {\it parallelizable manifold} is the one for which there exists a torsion which makes the manifold flat, i.e. makes its Riemann curvature tensor vanish. A necessary condition for parallelizability is that the torsion tensor should be covariantly constant,  
\be \nabla_\mu H_{\a\b\nu} = 0 \ ,\label{parallel}\ee
or in other words the torsion must be {\it parallel}. Except for the flat spacetimes which are trivially parallelizable, the curved parallelizable manifolds are  generically equipped with torsion. They arise naturally in most string theories as non-dilatonic NS-NS backgrounds of type II supergravity \cite{sadri}. Explicitly, we decompose the asymmetric connection into the Levi Civita connection and the contortion tensor
\be \hat\Gamma{^\mu}_{\nu\a} = {\Gamma^\mu}_{\nu\a} + {K^\mu}_{\nu\a} \ . \ee
%Contortion tensor 
%\be {K_{\mu\nu}}^\alpha = - {Q_{\mu\nu}}^\alpha + {Q_\nu}{^\alpha}_\mu - {Q^\alpha}_{\mu\nu} \Rightarrow {K_{\mu\nu}}^\alpha = - {K_\mu}{^\alpha}_\nu \label{contort}\ee
The Christoffel connection is symmetric in 2nd and 3rd indices while the contortion is antisymmetric between the two. 
%Thus, the torsion plays the similar role as the Christoffel connection. In addition, it is covariantly constant, \be \nabla T_{\mu\nu\a} = 0\ee
The generalized curvature tensor is expressed as a sum of the Riemann tensor (which depends only on the metric) and  the {\it torsional curvature}
\be \hat{R}_{\a\b\mu\nu} = R_{\a\b\mu\nu} + \nabla_\mu K_{\a\b\nu} - \nabla_\nu K_{\a\b\mu} + K_{\a\rho\mu} {K^\rho}_{\b\nu} - K_{\a\rho\nu} {K^\rho}_{\b\mu} \ .\ee 
Here $\nabla$ is the torsion-free covariant derivative. The parallelizability condition is $\hat{R}_{\a\b\mu\nu} = 0$. The generalized Ricci tensor is asymmetric 
\be \hat R_{\b\nu} = \hat R{^\a}_{\b\a\nu} = R_{\b\nu} +  \nabla_\a {K^\a}_{\b\nu} - \nabla_\nu {K^\a}_{\b\a} + {K^\a}_{\rho\a} {K^\rho}_{\b\nu} - {K^\a}_{\b\rho} {K^\rho}_{\a\nu} \ .   \ee 
The scalar curvature in presence of torsion is 
\be \hat R = R - \nabla^\b {K^\a}_{\b\a}   - K_{\a\b\rho} K^{\rho\a\b} \ .\ee
 If the Ricci tensor $\hat R_{\b\nu}$ is vanishing then the manifold is called {\it Ricci-parallelizable}.  In case of totally antisymmetric torsion like NS-NS field strength $H_{\mu\nu\a}$, the contortion tensor is equal to the torsion tensor,  $K_{\mu\nu\a} = T_{\mu\nu\a} = \frac{1}{2}H_{\mu\nu\a}$. Then the Ricci-parallelizability condition becomes
 \be \hat R_{\b\nu} = 0 = R_{\b\nu} + \dfrac{1}{2} \nabla_\a {H^\a}_{\b\nu} - \frac{1}{4}{H^\a}_{\b\rho} {H^\rho}_{\a\nu} \ .  \label{ricci-parallel}  \ee
For this, the symmetric and antisymmetric parts in eq. \ref{ricci-parallel} must vanish, which gives
\bea  R_{\b\nu} &=& \frac{1}{4} {H^\a}_{\b\rho} {H^\rho}_{\a\nu} \n\\ \textrm{and   }
\nabla_\a {H^\a}_{\b\nu} &=& 0 \label{ricci-parallel2}\ .
  \eea
Also the scalar curvature is $R = \dfrac{1}{4} H_{\a\b\rho} H^{\a\b\rho}$. All parallelizable manifolds are obviously also Ricci-parallelizable. The supergravity equations of motion for the $g_{\mu\nu}$ and $B_{\mu\nu}$ fields are the same as the Ricci-parallelizability condition \ref{ricci-parallel2}. 

\section{$D_p$-branes in Parallelizable NS-NS backgrounds}  \label{sec2}
Consider the propagation of a fundamental open string ending on a $D_p$-brane in a non-dilatonic NS-NS background, with the RR fluxes switched off. With the background metric $g_{\mu\nu}(x)$ and the two-form field $B_{\mu\nu}(x)$, the nonlinear sigma model action describing the string propagation is 
\be S = \dfrac{1}{4 \pi\a'} \int_{\Sigma} g_{\mu\nu}(X)\, dX^\mu \wedge *dX^\nu + \dfrac{i}{4 \pi\a'} \int_{\Sigma} (B_{\mu\nu}(X) + F_{\mu\nu}(X))\, dX^\mu \wedge dX^\nu \ , \label{nlsm-wedge} \ee
where $\Sigma$ is the string worldsheet, $x^\mu$ are the spacetime coordinates, $X^\mu(\tau,\sigma)$ are the string coordinates and $F = dA$ is the Maxwell field strength.    

\sp\ni
In this article, we will focus on weak string coupling and hence $\Sigma$ is assumed to have the topology of a disk. It is advantageous to use the Riemann normal coordinates (RNC) at the origin $x^\a = 0$ for the background fields in action \ref{nlsm-wedge}. In terms of RNC, the Taylor expansion of a tensor around the origin $x^\a = 0$ is simply given in terms of covariant tensors computed at the origin. Thus upto 2nd order, the normal coordinate expansions of the bulk fields are \cite{cornalba}
  \bea g_{\mu\nu}(x) &=& g_{\mu\nu}(0) - \frac{1}{3} R_{\mu\a\nu\b}\, x^\a x^\b + ..  \n\\ \textrm{and}\quad B_{\mu\nu}(x) &=& B_{\mu\nu}(0) +  \dfrac{1}{3} H_{\mu\nu\a} \,x^\a  + \frac{1}{4} \nabla_\b H_{\mu\nu\a}\, x^\a x^\b + .. \n\\
  \Rightarrow H_{\mu\nu\a}(x) &=& H_{\mu\nu\a}(0) \ \, .
 \label{rnc-bulk}  \eea
  Substituting the expansions in \ref{rnc-bulk} into the action \ref{nlsm-wedge}, it is re-expressed as \be S = S_0 + S_1 + S_2 + .. \ ,\ee in which $S_n$ is $n+2$ order in the  coordinate field $X^\mu$. Assuming weakly curved background, terms only upto the {\it leading order} in $H$ field contribute, where 
 \bea S_0 &=& \dfrac{1}{4 \pi\a'}\, g_{\mu\nu} \int_{\Sigma}  dX^\mu \wedge *dX^\nu + \dfrac{i}{4 \pi\a'} \int_{\Sigma} (B_{\mu\nu} + F_{\mu\nu}(X))\, dX^\mu \wedge dX^\nu \n\\ \textrm{and}\quad S_1 &=& \dfrac{i}{12 \pi\a'}\,  H_{\mu\nu\a}\, \int_{\Sigma}  X^\a \, dX^\mu \wedge dX^\nu \ .\eea 
 We are mainly interested in studying the effects of $S_1$ which represents  a small curved deviation from the flat background. 
  
 \sp\ni
 In presence of torsion, the symplectic structure 
 \be{\cal F}_{\mu\nu}(x) = B_{\mu\nu}(0) + F_{\mu\nu}(x) \ee
  is replaced by the gauge invariant combination 
 \be{\cal \tilde{F}}_{\mu\nu}(x) = B_{\mu\nu}(x) + F_{\mu\nu}(x) =  {\cal F}_{\mu\nu}(x) +  \dfrac{1}{3} H_{\mu\nu\a} \,x^\a \ ,\label{H0}\ee
 %where the torsion $H = d{\cal \tilde{F}} = dB$ is evaluated at the origin $x = 0$.
 Also, the Poisson structure $\theta = {\cal F}^{-1}$ is replaced by noncommutative parameter $\tilde{\theta} = {\cal \tilde{F}}^{-1}$.  Often the $A_{\mu}$ field is gauged away and one has a simple case of constant symplectic structure ${\cal F}_{\mu\nu} = B_{\mu\nu}(0)$ or $B_{\mu\nu}^{(0)}$.  %Thus we have,  
 %\be{\cal \tilde{F}}_{\mu\nu}(x) = B_{\mu\nu}(x)  =  B_{\mu\nu}(0) +  \dfrac{1}{3} H_{\mu\nu\a}(0) \,x^\a \ .\ee 
 The worldvolume deformation is now described by a nonassociative Kontsevich star product expansion \cite{cornalba}, 
\bea f \bullet g = f g + \dfrac{i}{2} \tilde{\theta}^{\a\b} (\partial_\a f)(\partial_\b g) - \dfrac{1}{8} \tilde{\theta}^{\a\mu} \tilde{\theta}^{\b\nu} (\partial_\a\partial_\b f) (\partial_\mu\partial_\nu g)\n\\ -\, \dfrac{1}{12} \tilde{\theta}^{\a\mu} \partial_\mu\tilde{\theta}^{\b\nu} \{(\partial_\a\partial_\b f)( \partial_\nu g) - (\partial_\b f) (\partial_\a\partial_\nu g)\} + O({\tilde{\theta}}^3)\ . \eea
  Thus in curved backgrounds, the brane worldvolume is not only deformed through the noncommutative parameter $\tilde{\theta}$, but also through a nonassociative parameter
  \be H = d{\cal \tilde{F}} =  dB \ . \ee 
  The associator of $\bullet$ product is defined by 
 \be (f \bullet g) \bullet h - f \bullet (g \bullet h) = \dfrac{1}{6}\tilde{\theta}^{\a\mu} \tilde{\theta}^{\b\nu} \tilde{\theta}^{\gamma\rho} H_{\mu\nu\rho} \,\partial_\a f \,\partial_\b g \,\partial_\gamma h \ , \ee
 which is non-zero since $H_{\mu\nu\rho}\neq 0$.  As derived in the Ref. \cite{cornalba}, after taking into account the corrections coming from the nonassociative Kontsevich star product, the effective open string metric $\tilde{G}$ in presence of NS-NS torsion is related to the flat space open string metric $G$ (in presence of constant two form background), as follows
 %\be \tilde{\theta}^{\a\b} =  \theta^{\a\b} + \dfrac{1}{3} \theta^{\a\mu} \theta^{\b\nu} H_{\mu\nu\rho} x^\rho \ee 
 \be \tilde{G}^{\a\b} = G^{\a\b} - \dfrac{1}{3} \theta^{\a\mu} G^{\b\nu} H_{\mu\nu\rho} x^\rho + \dfrac{1}{3} G^{\a\mu} \theta^{\b\nu}  H_{\mu\nu\rho} x^\rho \ . \label{osm-torsion1}\ee
 In terms of the constant two form $B_{\a\b}^{(0)}$,
\be \tilde{G}^{\a\b} = G^{\a\b} - \dfrac{1}{3}(B^{-1}_{(0)})^{\a\mu} G^{\b\nu} H_{\mu\nu\rho} x^\rho + \dfrac{1}{3} G^{\a\mu}(B^{-1}_{(0)})^{\b\nu}  H_{\mu\nu\rho} x^\rho \ . \label{osm-torsion2} \ee   
In the effective open string metric expressions \ref{osm-torsion1} and \ref{osm-torsion2}, the algebra is ordinary. The eqn. \ref{osm-torsion2} can be combined with the flat space formula of Seiberg-Witten metric $G_{\a\mu}$ (derived in Ref.  \cite{seiberg}), which connects the open and closed string parameters.

\section[A heuristic derivation of open string metric]{A heuristic derivation of open string metric in weakly curved background} \label{sec3}
In this section, the author suggests a heuristic approach of obtaining the effective open string metric in presence of torsion. Consider a $D_p$-brane in a parallelizable NS-NS background, we promote the constant two-form in the flat space formula given by Sieberg-Witten to a dynamical two-form.
  \bea G_{\mu\nu} &=& g_{\mu\nu} - (2\pi\a')^2 B^{(0)}_{\mu\rho} g^{\rho\a} B^{(0)}_{\a\nu} \n\\  \longrightarrow G^{(H)}_{\mu\nu} &=& g_{\mu\nu}(x) - (2\pi\a')^2 B_{\mu\rho}(x) g^{\rho\a} B_{\a\nu}(x) \eea
Assuming weakly curved background, we Taylor expand the bulk fields upto 1st order in coordinates $x^\a$. Using the expressions in eqn. \ref{rnc-bulk}, 
\bea G^{(H)}_{\mu\nu} &=& g_{\mu\nu} - (2\pi\a')^2 (B_{\mu\rho}^{(0)} +  \dfrac{1}{3} H_{\mu\rho\lambda} \,x^\lambda) g^{\rho\a} (B_{\a\nu}^{(0)} +  \dfrac{1}{3} H_{\a\nu\gamma} \,x^\gamma)\n\\ &=& g_{\mu\nu} - (2\pi\a')^2 B_{\mu\rho}^{(0)}g^{\rho\a} B_{\a\nu}^{(0)} - \, (2\pi\a')^2 \,\dfrac{1}{3} H_{\mu\rho\gamma} \,x^\gamma g^{\rho\a} B_{\a\nu}^{(0)} \n\\&&-\,  (2\pi\a')^2 \,\dfrac{1}{3} H_{\a\nu\gamma} \,x^\gamma g^{\rho\a} B_{\mu\rho}^{(0)}\n\\
\Rightarrow G^{(H)}_{\mu\nu} \textrm{ or } \tilde{G}_{\mu\nu} &=& G_{\mu\nu} + (2\pi\a')^2\, \dfrac{1}{3}\, (H_{\mu\rho\gamma} B_{\nu\a}^{(0)} + H_{\nu\a\gamma} B_{\mu\rho}^{(0)} ) \, x^\gamma g^{\rho\a}  \ .
 \label{richakapoor} \eea   
Eq. \ref{richakapoor} approximately describes the metric that the open strings see in presence of torsion, in a weakly curved parallelizable NS-NS background. Also note that on interchanging indices $\mu \leftrightarrow \nu$ in \ref{richakapoor}, we get \bea  \tilde{G}_{\nu\mu} &=& G_{\nu\mu} + (2\pi\a')^2\, \dfrac{1}{3}\, (H_{\nu\rho\gamma} B_{\mu\a}^{(0)} + H_{\mu\a\gamma} B_{\nu\rho}^{(0)} ) \, x^\gamma g^{\rho\a} \n\\
&\overset{\rho\leftrightarrow\alpha}{=}& G_{\nu\mu} + (2\pi\a')^2\, \dfrac{1}{3}\, (H_{\nu\a\gamma} B_{\mu\rho}^{(0)} + H_{\mu\rho\gamma} B_{\nu\a}^{(0)} ) \, x^\gamma g^{\a\rho} \n\\  &=&  \tilde{G}_{\mu\nu} \ .
 \eea
 Thus the emergent metric is symmetric. It is to note that the $x^\gamma$ in eq. \ref{richakapoor} are cartesian coordinates in terms of the polar. So for a $D_5$-brane with five spatial coordinates $(r, \b, \psi, \theta, \phi)$,  
\bea && x^t = x^0 = t \, ,\n\\ && x^r = x^1 = r\, \cos \b \, ,
\n\\ && x^\b = x^2 = r\, \sin\b \cos\psi \, , \n\\ && x^\psi = x^3 = r\, \sin\b \sin\psi \cos\theta \, , \n\\ && x^\theta = x^4 = r\, \sin\b \sin\psi \sin\theta \cos\phi \, , \n\\  \textrm{and } && x^\phi = x^5 = r\, \sin\b \sin\psi \sin\theta \sin\phi \ .
 \eea
\section{Deformation of $D_5$-brane in a parallelizable NS-NS background}\label{sec4}
Consider a $D_5$-brane in a parallelizable NS-NS background with the two-form equation of motion  
\bea
&&{\nabla}_{\a}H^{\a\mu\nu}\ =\ 0\nonumber\\ 
\Rightarrow &&\partial_{\a}H^{\a\mu\nu}\ +\ \dfrac{1}{2}\left( g^{\lambda\rho}\partial_{\a}\ g_{\lambda\rho}\right)
H^{\a\mu\nu}\ =\ 0\ \label{covar_torsion} , 
\eea
where $g_{\lambda\rho}$ is the flat worldvolume metric. Metric signature chosen is ``mostly plus". The $D_5$-brane is assumed to be a spherical brane. The 5-sphere on which the brane wraps is parametrized by $ r, \b, \psi, \theta$ and $\phi$.
 The line element over the $D_5$-brane is given in terms of worldvolume coordinates as follows, 
\be 
ds^2 = - dt^2 + dr^2 + \frac{r^2}{\kappa} \,\left(d\beta^2 + S^2_\b d\psi^2 + S^2_{\beta} S^2_\psi d\theta^2 + S^2_\b S^2_\psi S^2_\theta d\phi^2 \right)\ ,\ee
where $\kappa = 2\pi\a'$, $S_\beta\ =\sin\beta$, $S_\psi\ =\sin\psi$, $S_\theta\ =\sin\theta$ and $S_\phi\ =\sin\phi$ .  
The author found four exact solutions to eqn. \ref{covar_torsion} in 6 dimensions for the $D_5$-brane, assuming only one non-zero dynamical component of $B$-field for simplicity. There can be other solutions too, but four of them are:
 \bea 1. \qquad &&H_{\b\psi\theta} =  \frac{1}{\sqrt{\kappa}} \partial_\b B_{\psi\theta} = - \dfrac{m}{\kappa^{3/2}}\dfrac{sin\b}{sin\theta} \n\\ &&\Rightarrow B_{\psi\theta} = \dfrac{m}{\kappa} \dfrac{cos\b}{sin\theta} \\   \n\\
 2. \qquad &&H_{\b\theta\phi} =  \frac{1}{\sqrt{\kappa}} \partial_\b B_{\theta\phi} = - \dfrac{m}{\kappa^{3/2}} sin\theta\, sin\b \n\\ &&\Rightarrow B_{\theta\phi} = \dfrac{m}{\kappa}{sin\theta}\,{cos\b} \\   \n\\
  3. \qquad &&H_{\phi\b\theta} =  \frac{1}{\sqrt{\kappa}} \partial_\theta B_{\phi\b} =  \dfrac{m}{\kappa^{3/2}}sin\theta\, sin\b \n\\ &&\Rightarrow B_{\b\phi} = \dfrac{m}{\kappa} \cos\theta \sin\b \label{sol4}\\   \n\\
  4. \qquad &&H_{\b\psi\phi} =  \frac{1}{\sqrt{\kappa}} \partial_\b B_{\psi\theta} = - \dfrac{m}{\kappa^{3/2}} sin\b \n\\ &&\Rightarrow B_{\psi\phi} = \dfrac{m}{\kappa} cos\b   \eea  
  In this section, we choose the  solution \ref{sol4} to compute the emergent metric on the spherical $D_5$-brane. For that, we also assume the non-zero spatial components of constant $B$-field to be $B_{\b r}^{(0)} = B_{\theta r}^{(0)} = \dfrac{l}{\kappa}$. Here $l$ and $m$ are dimensionless constants, and the mass dimensions $[\kappa] = -2$, $[H] = 2$ and $[B] = 3$. The equation of motion for \ref{sol4} expands as follows
  \be \partial_{\b}H^{\b\mu\nu}\, + \partial_{\theta}H^{\theta\mu\nu}\, +\,\dfrac{1}{2} \left( g^{\lambda\rho}\partial_{\b}\ g_{\lambda\rho}\right)
H^{\b\mu\nu} +\, \dfrac{1}{2}\left( g^{\lambda\rho}\partial_{\theta}\ g_{\lambda\rho}\right) H^{\theta\mu\nu}= 0  \label{eom-rk1}
\ee 
One can easily verify that for $\mu, \nu = \theta, \phi$ in eq. \ref{eom-rk1}, the $\b$ equation is satisfied using the solution \ref{sol4} and for $\mu, \nu =  \phi, \b$, the $\theta$ equation is satisfied. 
Substituting the solution \ref{sol4} into the effective open string metric \ref{richakapoor} suggested by author, we get the emergent metric on the brane. It is given by
 \be d\tilde{s}^2 = \tilde{G}_{\mu\nu} dx^\mu dx^\nu \ . \label{emergentline}\ee
  In the expression \ref{emergentline}, $x^\mu$ are the polar coordinates $ t, r,\sqrt{\kappa} \b, \sqrt{\kappa} \psi, \sqrt{\kappa} \theta,\sqrt{\kappa}  \phi$.
 \ni 
   The line element on the curved $D_5$-brane is determined to be 
 \bea d\tilde{s}^2  &=& - dt^2 + \left\lbrace 1 +  \dfrac{\kappa\, l^2}{r^2}\left(1 + \dfrac{1}{S_\b^2 S_\psi^2}\right) \right\rbrace dr^2 - 2 \, \dfrac{m l \kappa  }{3 r }\,\dfrac{S_\theta^2 S_\phi}{S_\psi} \, dr d\b +\, 2 \, \dfrac{m l \kappa  }{3 r}  \, S_\b^2  S_\psi S_\theta^2 S_\phi\, dr d\theta  \n\\ &&
+\,2 \, \dfrac{m l \kappa }{3 r} \, \left\lbrace \dfrac{S_\theta C_\psi }{S_\psi^2} - S_\b^2  S_\psi S_\theta^2 C_\phi\right\rbrace\, dr d\phi   
+\left( \dfrac{r^2}{\kappa} + l^2\right) \kappa d\b^2 \,+\, 2\, l^2 \kappa d\b d\theta \,+\, r^2 S_\b^2\, d\psi^2 \n\\ &&+ \left(\dfrac{r^2}{\kappa}\, S_\b^2 S_\psi^2 + l^2\right)\kappa d\theta^2 \,+\, r^2 S_\b^2 S_\psi^2 S_\theta^2\, d\phi^2 \label{emergent} \ .
  \eea
  This metric appears to be singular at $r=0$ but on redefining the coordinate,
  \bea u = \dfrac{\sqrt{\kappa}\, l}{r} \Rightarrow dr = - \dfrac{\sqrt{\kappa}\, l}{u^2} \, du  \ \eea
and the mass dimension $[u] = 0$, the line element becomes \bea d\tilde{s}^2(u) &=& - dt^2 +  \dfrac{\kappa\, l^2}{u^4} \left\lbrace 1 +  u^2\left(1 + \dfrac{1}{S_\b^2 S_\psi^2}\right) \right\rbrace du^2 +  \dfrac{2\, m l \kappa  }{3  u }\,\dfrac{S_\theta^2 S_\phi}{S_\psi} \, du\, d\b \n\\ &&-\,  \dfrac{2\, m l \kappa}{3 u}  \, S_\b^2  S_\psi S_\theta^2 S_\phi\, du\, d\theta 
 -\,  \dfrac{2\, m l \kappa }{3 u}  \left\lbrace \dfrac{S_\theta C_\psi }{S_\psi^2} - S_\b^2  S_\psi S_\theta^2 C_\phi\right\rbrace du\, d\phi 
 \n\\ &&+\, l^2 \kappa\left( \dfrac{1}{u^2} + 1\right) d\b^2 \,+\, 2\, l^2\kappa \,d\b d\theta \,+\, \dfrac{l^2\kappa}{u^2}\, S_\b^2\, d\psi^2 \n\\ &&+ \,l^2\kappa\left(1+\dfrac{S_\b^2 S_\psi^2}{u^2} \right) d\theta^2 \,+\, \dfrac{l^2 \kappa}{u^2}\, S_\b^2 S_\psi^2 S_\theta^2\, d\phi^2 \label{emergent-u} \ .\eea 
 Now at $u = \infty$, the emergent metric is non-singular. This coordinate singularity implies that there is an event horizon in the metric \ref{emergent} and thus the geometry $d\tilde{s}^2(r)$ is that of a black 5-brane. The event horizon is located at the radial position where the radial component of the metric becomes singular, i.e. at 
 \bea &&\tilde{G}_{rr} =  \infty \n\\  &\Rightarrow& \dfrac{\kappa\, l^2}{r^2}\left(1 + \dfrac{1}{S_\b^2 S_\psi^2}\right)= \infty \n\\ &\Rightarrow& r = 0 \ .\eea
 Since the position of the singularity and the event horizon coincide, thus $r = 0$ is a ``null" singularity and the emergent geometry is that of an \textit{extremal} black 5-brane. Thus a spherical $D_5$-brane with torsion acts like a black 5-brane when its radius $r\rightarrow 0$. This is a useful case of gauge/gravity duality with an intractable non-commutative, non-associative gauge theory on a $D_p$-brane with torsion that is equivalent to a tractable, ordinary gravity theory on a black p-brane.  The \textit{near horizon geometry} is obtained in the large $u$ limit and is non-singular, 
 \bea d\tilde{s}^2(u) &=& -\, dt^2 +  \dfrac{\kappa\, l^2}{u^2} \left(1 + \dfrac{1}{S_\b^2 S_\psi^2}\right) du^2 + \kappa l^2 (d\b +  d\theta)^2 +  \dfrac{\kappa l^2}{u^2}\, d\Omega_4^2 \n\\ && +\,  \dfrac{2\, m l \kappa }{3  u } \left\lbrace\dfrac{S_\theta^2 S_\phi}{S_\psi} \, du\, d\b -  S_\b^2  S_\psi S_\theta^2 S_\phi\, du\, d\theta 
 -  \left( \dfrac{S_\theta C_\psi }{S_\psi^2} - S_\b^2  S_\psi S_\theta^2 C_\phi\right) du\, d\phi  \right\rbrace
  \label{emergent-nh} .\eea 
  Here, the term in curly bracket with factor $m$, is the torsion contribution. It appears that with some other choice of parallelizable NS-NS background, we can obtain familiar geometries with useful physical interpretations. %Further work needs to be done in this direction. %The future works of the author aim in this direction.
\section{Conclusion} 
In this paper, I have suggested a possible form of the open string metric in presence of torsion, assuming a weakly curved  background. This form is different from the open string metric given in my earlier works in a collaboration \cite{richa}. We use this metric to compute the emergent geometry  over a spherical $D_5$-brane. It turns out that the brane has a coordinate singularity at $r = 0$, and it thus acts like a black 5-brane for radius $r\rightarrow 0$. This is a special case of gauge/gravity duality with a non-commutative, non-associative gauge theory on the $D_5$-brane, which is dual to a simpler gravity theory with ordinary algebra on the black brane. 
%We obtain the \textit{near-horizon geometry} of the brane. 

%%%%%%%%%%%%%%%%%%%%%%%%%%
\section*{Acknowledgments}
%%%%%%%%%%%%%%%%%%%%%%%%%%
The author thanks Prof. Amitabha Mukherjee  for comments.

%**********************************************************%
\def\anp{Ann. of Phys.}
\def\cmp{Comm.Math.Phys. }
\def\prl{Phys.Rev.Lett. }
\def\prd#1{{Phys.Rev.} {\bf D#1}}
\def\jhep{JHEP\ {}}{}
\def\jaat{J.Astrophys.Aerosp.Technol.\ {}} {}
\def\cqg#1{{Class.\& Quant.Grav.}}
\def\plb#1{{Phys. Lett.} {\bf B#1}}
\def\npb#1{{Nucl. Phys.} {\bf B#1}}
\def\mpl#1{{Mod. Phys. Lett} {\bf A#1}}
\def\ijmpa#1{{Int.J.Mod.Phys.}{\bf A#1}}
\def\mpla#1{{Mod.Phys.Lett.}{\bf A#1}}
\def\rmp#1{{Rev. Mod. Phys.} {\bf 68#1}}
\def\ptep{Prog. of Theo.\& Exper.Phys.}
\def\jcap{J.Cosmo.\& Astropar.Phys.}
%**********************************************************%

\end{document}